# Spectrally Narrowed Edge Emission from Organic Light-Emitting Diodes: Evidence for Amplified Spontaneous Emission and Mirrorless Lasing

Yun Tian,[1] Zhengqing Gan,[1] Zhaoqun Zhou,[1] Ji-hun Kang,[2] Q-Han Park,[2] David W. Lynch,[1] Joseph Shinar[1]*

[1]Ames Laboratory-USDOE and Department of Physics and Astronomy, Iowa State University, Ames, IA 50011, USA.

[2]Department of Physics, Korea University, Seoul 136-701, Korea.

## Abstract

π-Conjugated materials, including small molecules and polymers, are attracting substantial attention as novel gain media in semiconductor lasers; they offer many potential advantages not achievable with conventional inorganic semiconductors: simple processing, low cost, easy tuneability of the spectrum, and large-area integration on flexible substrates. Optically pumped lasing action in various small molecular and polymeric π-conjugated materials has been demonstrated using several resonator configurations. However, electrically pumped organic semiconductor lasers, i.e., organic injection or diode lasers, remain elusive, presumably due to various loss mechanisms, e.g., charge (polaron)-induced absorption and metal electrode absorption. Here we report on evidence for amplified spontaneous emission (ASE), also known as mirrorless lasing (i.e., wherein some of the spontaneously emitted photons are amplified by stimulated emission during their propagation) in DC-driven small molecular organic light-emitting diodes (SMOLEDs). The evidence includes a dramatic spectral line narrowing, with a full width at half maximum (FWHM) of only 5 - 10 nm, and optical gain, of the edge-emission from SMOLEDs at room temperature. However, there is no clear indication of threshold behavior associated with this spectral narrowing. Nevertheless, this discovery should pave the way towards the realization of an organic diode laser.

---

*To whom correspondence should be addressed. E-mail: jshinar@iastate.edu

## I. INTRODUCTION

Unlike inorganic compound semiconductors, π-conjugated materials, which are organic semiconductors, are a four-level system [1], rendering them very desirable for application as a laser medium. The successful demonstration of optically pumped organic lasers has inspired intense efforts to achieve lasing via electrical injection in OLEDs [2–7]. This goal, however, has remained elusive. Under optical pumping, a singlet exciton (SE) is directly generated on a single molecule or conjugated segment of a polymer. The main optical loss is due to the propagation dissipation in organic semiconductor waveguides. Electrical excitation is a different process, where the electron and hole are injected separately into the active region to recombine to either SEs or triplet excitons (TEs). Strong optical losses arise from the presence of absorptive species such as polarons [8,9] and the incorporation of absorptive metal electrodes [10]. Moreover, the electroluminescence (EL) quantum efficiency is reduced substantially at high current density due to quenching of SEs by other SEs, by TEs, and by polarons [11,12]. This presents a major obstacle in reaching the lasing threshold. Yet this letter demonstrates edge emission from small molecular OLEDs (SMOLEDs) with key attributes of ASE, also termed mirrorless lasing: a drastically narrowed emission spectrum, and optical gain; however, no threshold bias or current for this emission is observed. The letter demonstrates that this spectrally narrowed edge emission (SNEE) results from an irregular waveguide mode propagating in the lateral direction along the OLED/glass substrate interface, which is quite different from the optically pumped ASE that results from regular waveguide modes confined within the organic layers.

## II. DEVICE PREPARATION AND EXPERIMENTAL METHODS

Various multilayer small molecular OLEDs, typically 21×21 matrix arrays of circular ~1.5 mm diameter pixels, were fabricated on indium tin oxide (ITO)-coated glass substrates. The emitting layer was N,N'-diphenyl-N,N'-bis(1-naphthylphenyl)-1,1'-biphenyl-4,4'-diamine (NPD), tris(quinolinolate) Al (Alq$_3$), or 4,4'-bis(2,2'-diphenylvinyl)-1,1'-biphenyl (DPVBi) (see Fig. 1). The $R_\square \sim 20$ /□, 140 nm-thick ITO coated 2"×2" glass substrates were treated in a UV/ozone oven to increase the ITO work function and facilitate hole injection. The substrates were then cleaned by detergent and organic solvents. The organic layers, CsF buffer layer, and Al cathode were deposited by thermal evaporation in a high-vacuum chamber (< 5 × 10$^{-6}$ Torr) installed in a glove box filled with Ar. By using a combinatorial sliding shutter technique, the OLEDs' thickness was varied systematically on the same batch. The Al cathode was deposited through a shadow mask containing 21×21 1.5 mm-diameter circular holes, yielding a 21×21 matrix array of OLED pixels.

The EL and photoluminescence (PL) surface and edge emission spectra were collected through an optical fiber probe connected to an OceanOptics Chem-2000 spectrometer. For polarization-dependent measurements, a linear polarizer was placed in front of the probe. All optical measurements were performed in air at room temperature.

III. RESULTS AND DISCUSSION

Fig. 2 shows the surface and edge emission spectra from a [glass]/[ITO]/[10 nm copper phthalocyanine (CuPc)]/[$x$ nm NPD]/[30 nm bathocuproine (BCP)]/[1 nm CsF]/Al OLED for various $x$; the nonemissive BCP served as both an electron-transporting and hole-blocking layer; the CsF layer enhances electron injection [13]. Thus, the recombination zone was localized within the NPD layer, as confirmed by the surface emission spectra. In contrast, the edge emission spectra exhibited a very interesting trend: In devices with a 40 – 80 nm thick NPD layer, the edge emission spectrum was rather broad, red-shifted relative to that in the normal direction, and nearly purely transverse magnetic (TM)-polarized. Similar edge emission spectra from similar OLED structures were reported by Bulovic et al [14]. They attributed these spectra to weak microcavity effects in OLEDs due to the reflective metallic cathode and ITO anode. Nevertheless, in devices with an NPD layer thickness ≥ 80 nm, a narrow emission band with a FWHM of 7.5 nm emerges at 410 nm, i.e., significantly blue-shifted relative to the surface emission, and the broad emission collapses; the narrow band polarization is transverse electric (TE) with a degree of polarization > 0.90. Finally, the narrow emission band red-shifts with increasing NPD layer thickness.

Alq$_3$-based OLEDs exhibited similar behavior above the threshold thickness of ~100 nm.

Fig. 3 shows the surface and edge emission spectra of a DPVBi-based OLED. The vertical microcavity effects induced by the metal cathode are evident from the red-shift of the surface emission spectra, from 443 nm to 470 nm, as the DPVBi thickness increases from 76 to 121 nm.

As in the NPD-based OLEDs, the edge emission is dramatically narrower, with ~6.5 nm FWHM, and red shifts as well, from 429 to 443 nm. Since the ratio of the broad and narrow peak emission wavelengths remains constant, we conclude that the vertical cavity resonance, to some extent, may contribute to the edge emission as well. Moreover, the sharp edge emission peak is a linearly TE polarized mode with a degree of polarization as high as 0.94.

Fig. 4 shows the narrow edge emission from another DPVBi OLED, where FWHM ≈ 5 nm. We note that the FWHM was independent of the driving voltages, so, in particular, there was no apparent threshold voltage or current density above which the edge emission spectrum narrowed.

To recap, for the NPD or $Alq_3$ OLEDs, there is a critical thickness for the SNEE to occur, along with a polarization transition from a TM to a strong TE mode. However, no critical thickness was observed in DPVBi OLEDs, probably because the minimum thickness already exceeded the critical thickness.

The behavior of the SNEE indicates that it is related to a waveguiding effect. Since the refractive indices of the organic ($n \sim 1.7$) and ITO ($n \sim 1.9$) layers are greater than that of the glass ($n \sim 1.5$), the combined organic and ITO layers constitute a dielectric slab waveguide core, with the glass and Al as the waveguide claddings, resulting in a three-slab asymmetric waveguide structure. This structure preferentially supports the TE polarization [10]. It might be argued, however, that such high TE polarization arises because the planar DPVBi and NPD molecules lie preferentially in the plane of the substrate, so more light is emitted with a TE polarization due to the preferential orientation of the SE dipoles in the plane of the substrate [15]. Yet the SNEE from OLEDs based on spiro-DPVBi and $Alq_3$, which are intrinsically nonplanar, was also fully TE polarized. Additionally, in SMOLEDs fabricated by thermal evaporation, where the organic layers are amorphous, the dipoles are most likely randomly oriented.

To explore the nature of the SNEE, we begin by noting that the optical propagation loss in a metal-clad waveguide is primarily caused by the strong absorption of the evanescent tail of the optical field that penetrates the metal electrode [10,14]. Such a loss is estimated to be ~1000 $cm^{-1}$ [10,14]. In addition, the ITO is also lossy [16], particularly at $\lambda < 450$ nm. Consequently, the waveguided light would be completely absorbed before it contributes to any far-field edge emission (see below). Therefore the SNEE is probably not due solely to the regular waveguide modes, which are trapped in the organic and ITO layers, and cannot escape unless there is enough optical gain. This conclusion is supported by edge emission measurements on patterned

OLEDs in which the organic and ITO layers at the edge of the glass substrate were etched off and replaced by a tape, thus blocking the waveguide modes emerging from the faces of the organic and ITO edges: The resulting spectral shape, polarization and intensity were almost identical to those of the SNEE from OLEDs whose ITO layers remained unpatterned, indicating that the SNEE exits from the edge of the glass substrate and cannot be attributed to a regular waveguiding effect.

To clarify any geometric optics effects, a study of modes in the composite Al/organic/ITO/glass waveguide was conducted. We required that the wave in the ITO not be totally reflected at the organic interface, but that it be totally internally reflected at the ITO/glass interface. These conditions limited the range of propagation angles inside the waveguide. Phase shifts due to propagation and reflections were required to total to an integral multiple of $2\pi$ per repeat distance. For the thicknesses used, ca. 100 - 200 nm, many TE and TM modes were found, alternating in vacuum wavelength. Several of these modes were within the envelope of the luminescence spectrum of the organic. Losses occurred at each reflection; they were larger for TM modes, and upon propagation in the ITO when a complex refractive index $n = 1.90 - 0.04i$ was introduced [16]. However, these losses were not so great as to preclude propagation distances $d \sim 100$ μm without optical gain. Yet for $d \geq 1$ mm, some gain is required. However, this study could not determine whether the edge emission was due to a waveguide mode itself and/or to an evanescent wave in the glass. By decreasing the angle of incidence by 0.6° (1.6°) below the critical ITO/glass interface angle, a "leaky waveguide" mode (so called because it leaks into the glass substrate) was simulated. The mode wavelength blue-shifted by 16 (28) nm, and a wave propagated in the glass at grazing angles of 5.3° (8.5°) from the interface. The transmission coefficients were 0.38 (0.54), leading to observable intensity from the edge of the glass, but significant loss from the waveguide mode.

Pauchard *et al.* described an irregular mode that is confined at the polymer/SiO$_2$ interface in a multilayer field-effect transistor structure [17], which showed a propagation loss coefficient of only 9 cm$^{-1}$, thus leading to a considerably reduced optical gain threshold compared to the regular guided mode. Recently, an optically-pumped irregular waveguide mode with a narrowed spectrum was identified and characterized through grazing angle edge emission measurements [18,19]. Such an irregular mode is generally identified with a leaky waveguide mode traveling at a grazing angle adjacent to the organic/glass interface, which emerges from the substrate edge nearly parallel to the plane of the organic layer. The origin of spectral narrowing of this mode is believed to be due to the interference between the multiple reflections within the organic layer [18], or between the multiple leaked beams along the organic/glass interface [19], quite similar to a Fabry-Perot-like microcavity behavior. Indeed, Li *et al.* observed a low gain threshold for optically-pumped grazing-angle edge emission from a three-slab asymmetric waveguide structure [20], which was attributed to a cavity enhancement effect. More recently, Blinov *et al.*[21] demonstrated that under certain conditions, leaky-mode lasing has a lower threshold than regular waveguide lasing, and the leaky-mode laser action suppresses the efficient regular waveguide laser action.

The observed SNEE is therefore most likely due to a leaky waveguide mode, or perhaps other unknown irregular waveguide modes, propagating along the ITO layer/glass substrate interface. Using a luminance-meter, the SNEE was observed to occupy a stripe-like bright area at the edge of the glass substrate whose position is consistent with the irregular waveguide modes.

As mentioned above, interference effects might also be responsible for the SNEE. Finite difference time domain (FDTD) simulations were performed to solve Maxwell's equations for the OLED waveguide structure; the results suggest that the waveguiding effect alone can induce SNEE with certain device sizes (see the Supplementary Information). The possibility that the SNEE can be attributed to ASE, where stimulated emission plays the dominant role, was also considered. The generally accepted approach to identify ASE is the variable stripe length (VSL) method, i.e., the edge emission is measured from stripe-shaped OLEDs of variable length $l$ [22]. Then ASE mandates that the spectra should be broad at short $l$ and become narrower with increasing $l$, and that the output intensity versus $l$ should exhibit supralinear behavior [22]. In contrast, if a mechanism other than ASE dominates, the emission spectrum should not be affected by $l$ and the intensity should increase linearly or sublinearly with $l$.

Figure 5a shows the normalized edge emission spectra of OLED stripes with $1 \leq l \leq 12$ mm (and a fixed 1 mm stripe width) at $V \approx 10$ V and $J \approx 1$ mA/cm$^2$. Note the weakness of the TM mode peaks at short wavelength relative to the TE mode peaks with increasing $l$.

Figure 5b shows the TE mode peak edge emission intensity vs. $l$. We fit the data by the well-known ASE equation $I = (AI_p/g)[\exp(gl) - 1]$, where $I_p$ is the pump intensity, $g$ is the net gain coefficient, and $l$ is the excitation length. The excellent match of the data points with the fitting curve is evident. Above certain excitation stripe length, the intensity is expected to level off due to a saturation effect [22], Fig. 5b exhibits such a saturation clearly for $l > 9$ mm.

Figure 5c shows the dependence of the FWHM of the TE mode edge emission spectra on $l$. Note that the decreasing FWHM approaches an asymptotic value of ~14 nm at long $l$ ($l > 9$ mm), this further confirms the onset of a saturation effect due to the depletion of excitation density by stimulated emission [22].

In conclusion, the observed behavior is in good agreement with ASE and suggests that the SNEE cannot be attributed to interference or waveguiding effects only. The results also rule out the possibility of superfluorescence and biexcitonic spontaneous emission. Indeed, they provide strong evidence of light amplification and optical gain due to ASE. Note that an excitation length range of 0.1 - 1 mm is commonly used to study ASE in optically pumped organic waveguides. Therefore the optical gain coefficient derived from Figure 5b is lower than that in optically pumped organic structures by at least an order of magnitude.

Although the results indicate that ASE can be achieved in SMOLEDs, which is very significant for the development of organic diode lasers, the nature of the leaky waveguide mode in terms of its propagating pathway and characteristics is still not clear. To explore its nature further, the propagation loss of this mode was estimated by measuring the peak edge emission intensity vs the distance between the OLED pixel and the edge of the glass substrate. It was found that the propagation loss coefficient is nearly zero, indicating that this mode undergoes almost no attenuation during its propagation before exiting from the substrate edge. Such an unusual feature renders the mode very distinct from the regular waveguide modes whose propagation loss coefficient can be as high as 40 cm$^{-1}$. Although other leaky waveguide modes may share some common attributes with the mode identified in this letter, the propagation loss coefficient of ~8 cm$^{-1}$ associated with this leaky mode suggests that they are different in nature due to different mode field distributions and spatial confinement within the OLED structure. The exceptionally low threshold for ASE observed in this study may be a consequence of the extremely low propagation loss.

**Acknowledgements**

Ames Laboratory is operated by Iowa State University for the US Department of Energy (USDOE) under Contract No. W-7405-ENG-82. The work in Ames was supported by the

Director for Energy Research, Office of Basic Energy Sciences, USDOE. The work in Korea was supported in part by q-Psi/KOSEF and the Seoul R&BD Program.

**References**

1. M. D. McGehee, A. J. Heeger, Adv. Mater. **12**, 1655 (2000).
2. N. Tessler, G. J. Denton, R. H. Friend, Nature **382**, 695 (1996).
3. F. Hide *et al*., Science **273**, 1833 (1996).
4. M. Berggren, A. Dodabalapur, R. E. Slusher, Z. Bao, Nature **389**, 466 (1997).
5. V. G. Kozlov, V. Bulovic´, P. E. Burrows, S. R. Forrest, Nature **389**, 362 (1997).
6. V. Bulovic´, V. G. Kozlov, V. B. Khalfin, S. R. Forrest, Science **279**, 553 (1998).
7. M. D. McGehee *et al*., Appl. Phys. Lett. **72**, 1536 (1998).
8. V. G. Kozlov *et al*., Appl. Phys. Lett. **74**, 1057 (1999).
9. N. Tessler *et al*., Appl. Phys. Lett. **74**, 2764 (1999).
10. N. Tessler, Adv. Mater. **11**, 363 (1999).
11. M. A. Baldo, R. J. Holmes, S. R. Forrest, Phys. Rev. B **66**, 035321 (2002).
12. H. Yamamoto *et al*., Appl. Phys. Lett. **84**, 1401 (2004).
13. L. S. Hung *et al.*, Appl. Phys. Lett. **70**, 152 (1997); G. E. Jabbour *et al.*, Appl. Phys. Lett. **73**, 1185 (1998).
14. V. Bulovic´ *et al*., Phys. Rev. B **58**, 3730 (1998).
15. K. H. Yim, R. Friend, J. S. Kim, J. Chem. Phys. **124**, 184706 (2006).
16. J. S. Kim *et al*., J. Appl. Phys. **88**, 1073 (2000).
17. M. Pauchard *et al*., Appl. Phys. Lett. **83**, 4488 (2003).
18. T. Kawase *et al*., Synthetic Met. **111/112**, 583 (2000).
19. A. Penzkofer *et al*., Opt. Commun. **229**, 279 (2004).
20. F. Li *et al*., J. Appl. Phys. **99**, 013101 (2006).
21. L. M. Blinov *et al.*, Appl. Phys. Lett. **89**, 031114 (2006).
22. M. D. McGehee *et al.*, Phys. Rev. B **58**, 7035 (1998).

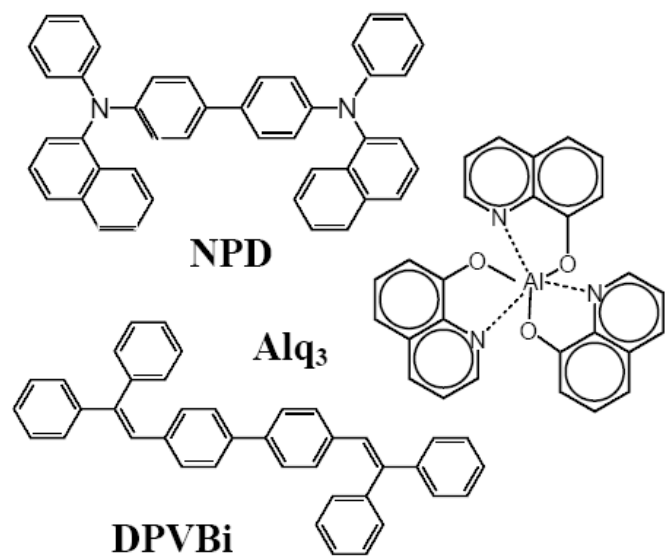

**Fig. 1.** Molecular structures of the small molecular weight organic materials used in this study.

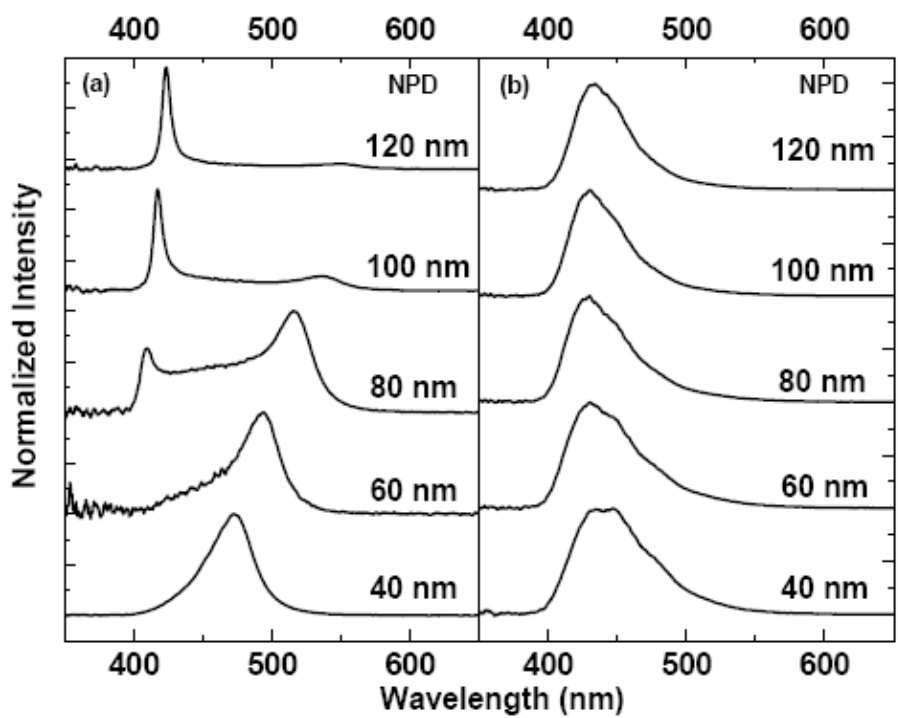

**Fig. 2.** (**a**) Edge emission spectra and (**b**) surface emission spectra from an OLED with the structure glass/ITO/[10 nm CuPc]/[$x$ nm NPD]/[30 nm BCP]/[1 nm CsF]/Al, with $x$ = 40, 60, 80, 100, 120 nm.

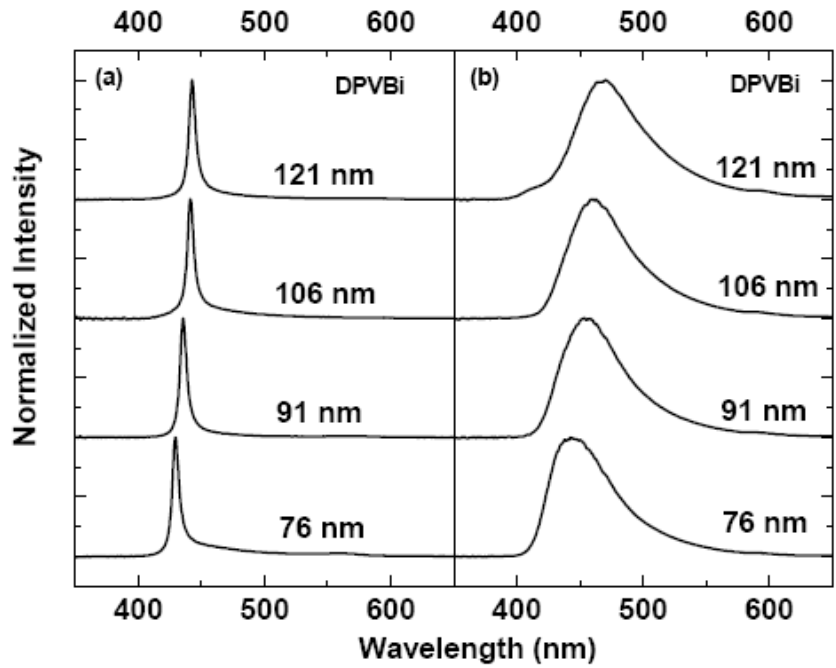

**Fig. 3.** (**a**) Edge emission spectra and (**b**) surface emission spectra from an OLED with the structure glass/ITO/[5 nm CuPc]/[46 nm NPD]/[$x$ nm DPVBi]/[6 nm Alq$_3$]/[1 nm CsF]/Al, with $x$ = 76, 91, 106 and 121 nm.

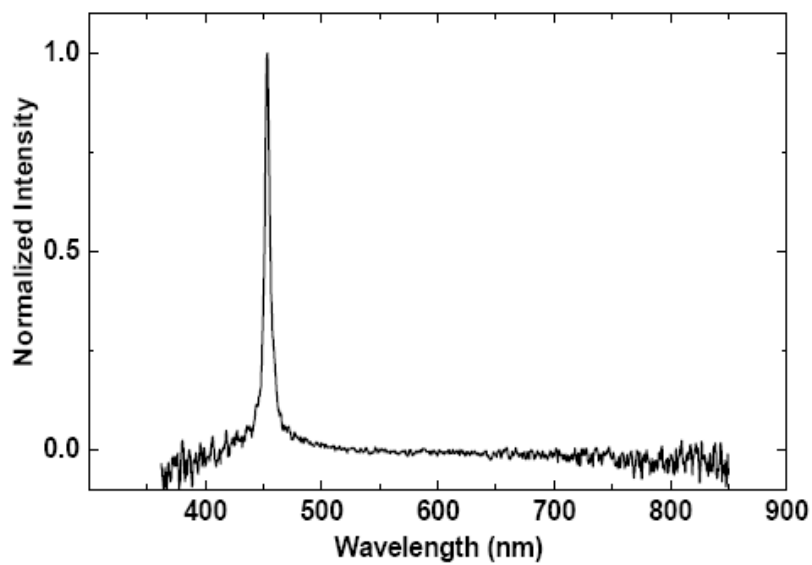

**Fig. 4.** Edge emission spectrum of an OLED with the device structure: glass/ITO/[5 nm CuPc]/[45 nm NPD]/[120 nm DPVBi]/[7 nm Alq$_3$]/[1 nm CsF]/Al.

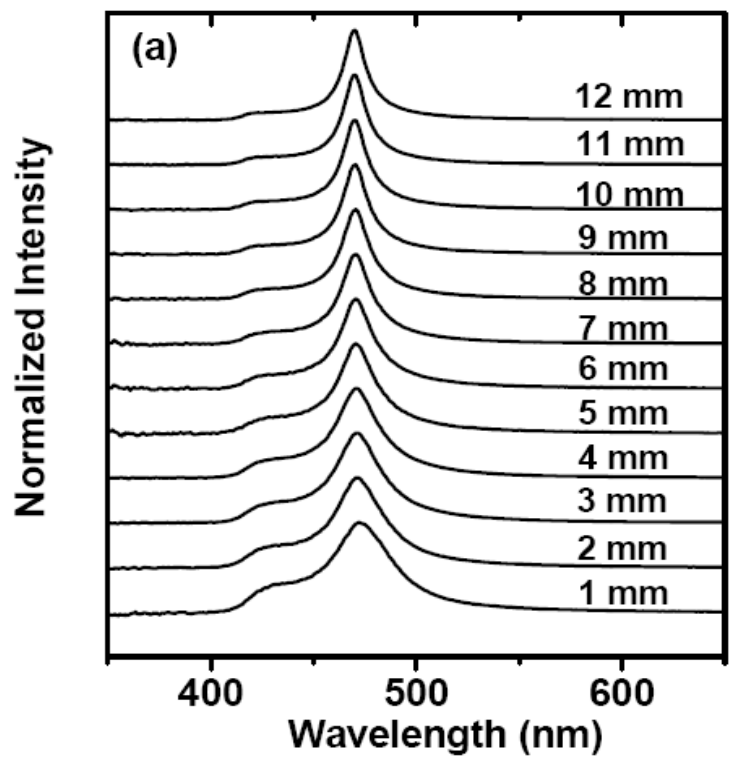

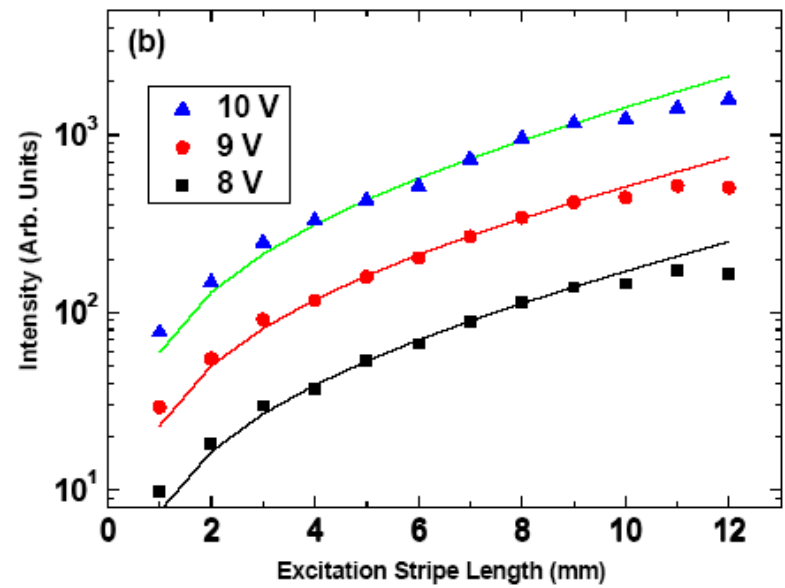

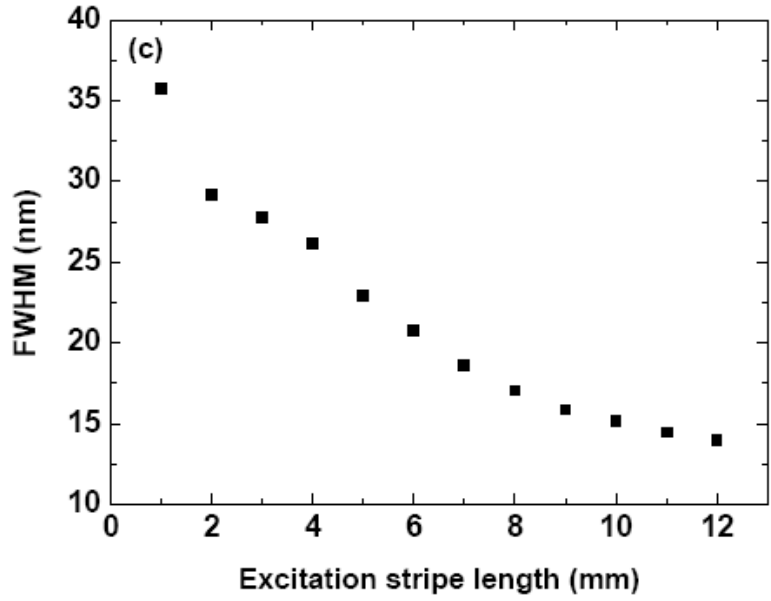

**Fig. 5.** (**a**) The edge emission spectra at the different stripe lengths. The OLED structure is glass/ITO/[40 nm NPD]/ [60 nm DPVBi]/[6 nm Alq$_3$]/[1 nm CsF]/Al. (**b**) The peak edge emission intensity *vs* the stripe length; the lines are the best fits of the function $y = (A/g)[\exp(gl) - 1]$. (**c**) The FWHM of the edge emission spectrum *vs* the stripe length.